\documentclass[12pt,onecolumn]{emulateapj}

\usepackage{epsf}
\usepackage{epsfig}

\def\wisk#1{\ifmmode{#1}\else{$#1$}\fi}
\def\lsim   {\wisk{_<\atop^{\sim}}}

\begin{document}

\title{Lensing and the Warm Hot Intergalactic Medium.}

\author{
F. Atrio-Barandela\altaffilmark{1},
J. P. M\"ucket\altaffilmark{2}}
\altaffiltext{1}{F{\'\i}sica Te\'orica, Universidad de Salamanca,
37008 Salamanca, Spain; atrio@usal.es}
\altaffiltext{2}{Leibniz-Institut f\"ur Astrophysik,
14482 Potsdam, Germany; jp\_muc@web.de}

\begin{abstract}
The correlation of weak lensing and Cosmic Microwave Anisotropy 
(CMB) data traces the pressure distribution of the hot, ionized gas 
and the underlying matter density field. The measured correlation is
dominated by baryons residing in halos. Detecting the contribution
from unbound gas by measuring the residual cross-correlation 
after masking all known halos requires a theoretical understanding 
of this correlation and its dependence with model parameters.
Our model assumes that the gas in filaments is well described
by a log-normal probability distribution function, with temperatures
$10^{5-7}$K and overdensities $\xi\le 100$. The lensing-comptonization 
cross-correlation is dominated by gas with overdensities in the 
range $\xi\approx[3-33]$; the signal is generated 
at redshifts $z\le 1$. If only 10\% of the measured cross-correlation
is due to unbound gas, then the most recent measurements set an upper limit
of $\bar{T}_e\lsim 10^6$K on the mean temperature of Inter Galactic Medium.
The amplitude is proportional to the baryon fraction stored in filaments. 
The lensing-comptonization power spectrum peaks at a different scale than 
the gas in halos making it possible to distinguish both contributions. 
To trace the distribution of the low density and low temperature 
plasma on cosmological scales, the effect of halos will have to be
subtracted from the data, requiring observations with larger signal-to-noise ratio
than currently available.
\end{abstract}


\section{Introduction.}

A recent account of the baryon distribution in the local Universe
concluded that about half the baryons synthesized in the Big-Bang
have yet to be identified (Shull, Smith \& Danforth 2012)
confirming an earlier deficit of baryons found by
Fukugita, Hogan \& Peebles (1998). 
Numerical simulations (Cen \& Ostriker 1999; Dav\'e et al 1999, 2001, 
Cen \& Ostriker 2006, Smith et al 2011) indicated that only 10-20\% of all 
baryons are in collapsed objects. Baryons in Intergalactic
Medium (IGM) exist in a wide range of densities and temperatures.
Penton, Stocke \& Shull (2004) and Lehner et al (2007) 
concluded that another $\sim$30\% resides in low redshift Ly$\alpha$ absorption 
systems, while the rest could reside in the shock-heated IGM with temperatures 
$10^{5-7}$K and overdensities $\xi\le 100$. This unbound gas is usually known 
as Warm-Hot Intergalactic Medium (WHIM). Identification of the WHIM phase and
its spatial distribution at low redshift is an on-going theoretical and 
observational effort (for a review, see McQuinn 2016). The low density 
makes it difficult to detect the WHIM in emission (Soltan 2006); more 
promising is using absorption lines in the far ultraviolet to the soft 
X-ray range but some earlier detections remain controversial (Shull et al 2012).
Cappelluti et al (2012) and Roncarelli et al (2012) have searched for the
contribution of the WHIM to the diffuse X-ray emission but failed to find
a statistically significant result.

Since the WHIM is highly ionized, there has been an extensive search on 
the thermal and kinematic Sunyaev-Zel'dovich CMB temperature anisotropies
(hereafter tSZ and kSZ, Sunyaev \& Zeldovich 1970; Sunyaev \& Zeldovich 1972)
generated by this baryon component (Atrio-Barandela \& M\"ucket 2006;
Atrio-Barandela et al 2008). Cross-correlation of CMB temperature data
from {\it WMAP} or {\it Planck} with matter templates produced only marginal evidence 
of tSZ anisotropies due to the WHIM (Suarez-Vel\'asquez et al 2013b, 
G\'enova-Santos et al 2013, G\'enova-Santos et al 2015).
Combining X-ray and tSZ observations could be a promising tool to study the WHIM
(Ursino, Galeazzi \& Huffenberger 2014).
The first evidence of warm-hot gas beyond the virial radius of clusters was
presented in Planck Collaboration (2013) who detected a filamentary structure
between the cluster pair A399-A401. The distribution of gas in a cosmic web has 
also been confirmed by XMM-Newton observations of the cluster Abell 2744 by
Eckert et al (2015) who found filamentary structures of gas at
temperature $10^7$K and coherent over a scale of 8Mpc. 
At those temperatures and densities, the kSZ effect can have a contribution
of similar amplitude to the tSZ effect. The kSZ 
effect has been used to trace large scale peculiar
velocity fields (Kashlinsky et al 2008, Atrio-Barandela et al 2015) 
and the anisotropies due to the pair-wise velocity dispersion of clusters 
and galaxies have been measured (Hand et al 2012, Soergel et al 2016, 
Schaan et al 2016, de Bernardis et al 2017). These latter observations probe 
baryons on cluster and galaxy scales but have not yet provide
a measurement of the fraction of free electrons.
A search of the kSZ anisotropies due to the WHIM found no statistically 
significant evidence in {\it WMAP} data (G\'enova-Santos et al 2009).
Only recently, Hern\'andez-Monteagudo et al (2015) and
Planck Collaboration (2016) have presented evidence of the 
peculiar motion of extended gas on Mpc scales with a statistical 
significance at the $3-3.7\sigma$ level. Hill et al (2016)
measured the kSZ effect correlating {\it WMAP} and {\it Planck} data with a
galaxy sample from the Wide-field Infrared Survey Explorer (WISE) verifying
that baryons approximately trace the Dark Matter (DM) distribution down
to $\sim$Mpc scales.

The cross-correlation of gravitational lensing maps with tSZ anisotropies
is another potential probe of the relation between the hot, ionized gas and
the matter density field. Hill \& Spergel (2014) determined the cross-power spectrum of 
weak lensing of the CMB with the tSZ anisotropies measured by {\it Planck} at
the $6\sigma$ confidence level, 
obtaining a constrain on the bias between the hydrostatic mass and the true mass of
clusters and groups at redshifts $z\le 2.5$. This authors  interpreted their
signal as being produced by baryons in halos. In parallel, 
Van Waerbeke, Hinshaw \& Murray (2014) found a 
detection of the cross-correlation between the tSZ signal from {\it Planck} and
the galaxy lensing convergence from the Canada-France-Hawaii Telescope 
Lensing Survey (CFHTLenS) with the same level of significance. 
Originally, the data were interpreted as the signal 
from warm and diffuse baryons. Since the distribution of galaxies in
the survey peaks at $z=0.37$, this result suggested that 
a large fraction of the missing baryon population had been identified.
New studies and numerical simulations demonstrated that the majority of 
the signal came from a small fraction of baryons within halos
(Ma et al 2015, Hojjati et al 2015, Battaglia, Hill \& Murray, 2015). 
On large angular scales the simulations showed a correlation slightly above that 
of the halo model prediction, pointing to a $10-15\%$ contribution from unbound
gas. The latter contribution is degenerate with respect to cosmological
and physical parameters and the data did not permit a robust inference 
(Battaglia et al 2015).  Hojjati et al (2016) improved the statistical
significance of the lensing -- tSZ cross-correlation by using a larger 
weak lensing map derived from the Red Sequence Cluster Lensing Survey (RCSLenS)
and found that their signal was best interpreted if AGN feedback
removed a large quantity of hot gas from galaxy groups.

To estimate the contribution of unbound gas to the
tSZ--lensing cross-correlation results described above requires
an analytical model that correctly predicts the amplitude and shape
of the expected signal. In Atrio-Barandela \& M\"ucket (2006) we 
described the unbound gas in the weakly non-linear filaments 
by means of the log-normal Probability Distribution Function (PDF). In this
article we use this description of the unbound gas to predict
the cross-correlation of the lensing convergence due to the large scale
matter distribution and the tSZ temperature anisotropies.
The outline of this paper is as follows: In Section~2 we 
describe the model and compute the tSZ--convergence cross-correlation; 
the derived expressions are solved numerically and the results are presented
in Section~3; finally, our conclusions are summarized in Section~4.

\section{Lensing-tSZ correlation in the filament model.}

The WHIM generates temperature anisotropies on the 
CMB via the tSZ. If $n_e$ and $T_e$ are
the electron density and temperature along the line of sight then
the anisotropy generated by the free electrons residing in the potential
wells of the WHIM filaments in units of the current CMB black-body temperature
$T_0$ is $\Delta T_{tSZ}/T_0=Y_CG(\nu)$. The comptonization 
parameter measures the integrated electron pressure along the line of sight, 
$Y_C=k_B\sigma_T/m_ec^2\int n_eT_e adw$ with $a$ the scale factor, $w$ the comoving
radial distance, $m_ec^2$ the electron annihilation temperature, $k_B$ the Boltzmann constant
and $\sigma_T$ the Thomson cross section; $G(\nu)=(x\coth(x/2)-4)$
gives the frequency dependence of the tSZ effect being $x=h\nu/k_BT_0$
the reduced frequency with $h$ the Planck constant and $\nu$ the frequency of observation.
This frequency dependence is different from that of any other known foreground
making the tSZ anisotropy possible to distinguish from other
CMB anisotropies given sufficient multi-frequency coverage. 
The data are usually expressed in terms of the comptonization 
parameter $Y_C$ instead of the temperature anisotropy.

The intrinsic CMB temperature anisotropies are lensed by
the large scale structure traced by galaxy catalogs.
The tSZ anisotropies are themselves generated by the ionized
gas within the same large scale structure.
The two-point correlation function of the lenses and the spatial variations of 
the electron pressure along the line of sight is the weighted average of the
lensing kernel $\Delta\kappa_{eff}$ due to the large scale structure 
traced by galaxy catalogs and the anisotropies
generated by the tSZ effect of the ionized gas. The weight is given by the
probability that the gravitational fields that lens the primary 
CMB anisotropies contain the electrons that generate the tSZ anisotropies.
If the ionized gas generates a comptonization 
parameter $\Delta Y_C=(k_B\sigma_T/m_ec^2)n_eT_ea(dw/dz)dz$ 
at redshift $z_1$ in the direction $\hat{x}_1$
and the lenses are located at $z_2$ in direction $\hat{x}_2$
then their correlation is 
\begin{equation}
C(\theta)\equiv
\langle\kappa_{eff} Y_C\rangle(\theta)= \int_0^{z_H}\int_0^{z_H}\langle \Delta Y_C
(\hat{x}_1,w_1)\Delta\kappa_{eff}(\hat{x}_2,w_2)\rangle dz_1 dz_2 ,
\label{eq:cfull}
\end{equation}
where $\theta$ is the angle between the directions
$\hat{x}_1$ and $\hat{x}_2$, i.e., $\cos\theta=\hat{x}_1\cdot\hat{x}_2$ and
$w_1,w_2$ are the comoving radial distances (for notation and definitions, 
see Bartelmann \& Schneider 2001). The integration extends out to
the redshift of the surface of the last scattering, $z_H$.
The average $\langle\cdots\rangle$ takes into account the 
distribution of the WHIM filaments and of the lenses and their correlation. 
Let us briefly summarize our WHIM model and the effect of a population
of lenses before discussing their statistics.

\subsection{The log-normal distribution of WHIM filaments.}

Numerical simulations have shown that at redshifts $z>1$ and at small 
scales the IGM forms filaments of mildly non-linear overdensities, giving rise to 
the observed Ly$\alpha$ forest. At $z<1$ most of the IGM baryons 
resides in shock-heated regions of low density gas at temperatures $0.01-1$KeV
(Shull et al 2012) and sizes larger than 1 Mpc (Cen \& Ostriker, 2006). We 
model the distribution of this unbound IGM gas as 
a log-normal random field evolving with time. The log-normal  PDF was 
introduced in Cosmology by Coles \& Jones (1991) 
to describe the non-linear distribution of matter 
in the Universe when the peculiar velocity field was still in the 
linear regime. Based on the improved Wiener density 
reconstruction from the Sloan Digital Sky Survey, Kitaura et al (2009) found 
that this distribution describes the statistics of the matter 
inhomogeneities on scales larger than $7h^{-1}$Mpc. In the log-normal 
approximation, the number density of baryons at $\vec{x}$, 
located at redshift $z$ and at a proper distance $|\vec{x}(z)|$
is $n_B(\vec{x},z)=n_0(z)\xi$,
where $\xi$ is a log-normal distributed random variable normalized to
have unit mean, $\langle\xi\rangle=1$, and
$n_0(z)=f_e\rho_B(1+z)^3/\mu_B m_p$ is the mean baryon number density,
$\rho_B$ the baryon density, $f_e$ the fraction of baryons in the WHIM,
$m_p$ the proton mass, $\mu_B=4/(8-5Y)$ the mean
molecular weight of the IGM and $Y$ the He fraction by weight that
we fixed to the value $Y=0.24$. 
The non-linear baryon density contrast $\xi$ in units
of the baryon mean density should not be confused with $\delta$
or $\delta_B$, respectively the matter and IGM baryon overdensities in the
{\it linear regime}; $\xi$ is given in terms of 
the gaussian distributed variable $\delta_B$
(Choudhury et al 2001, Atrio-Barandela \& M\"ucket 2006)
\begin{equation}
\xi={\rm e}^{\delta_B(\vec{x},z)-\sigma_B^2(z)/2} ,
\label{eq:logn}
\end{equation}
where $\sigma_B^2(z) = <\delta_B^2(\vec{x},z)>$ is the variance of the 
zero-mean linear IGM baryon density field.
The number density of electrons in the IGM, $n_e$, is obtained 
by assuming equilibrium between recombination and photo-ionization and 
collisional ionization. For the conditions of the IGM, temperatures in 
the range $10^5-10^7$K and density contrasts $\xi \le 100$, the gas can 
be considered fully ionized so $n_e\approx n_B$.

The spectrum of density fluctuations of the baryons in the IGM is
related to the DM density contrast $\delta_{DM}$ by (Fang et al 1993)
\begin{equation}
\delta_B(k,z)=\frac{\delta_{\rm DM}(k,z)}{[1+k^2L_{0}^2]} .
\label{pk}
\end{equation}
The cut-off length $L_{0}$ corresponds to the scale below 
which baryon density perturbations are smoothed due to physical processes.
The variance of the baryon density field is given by
\begin{equation}
\sigma_B^2(z)=\frac{D_+^2(z)}{2\pi^2}\int \frac{P_{DM}(k)}{[1+L_0^2(z)k^2]^2}k^2dk ,
\end{equation}
being $D_+(z)$ the linear growth factor of matter density perturbations.

\subsubsection{Baryon damping scales.}

At redshifts $z\le 1$ small scale baryon perturbations are 
erased by shock-heating (Klar \& M\"ucket 2010). If $T_{\mathrm{IGM}}$ is 
the mean IGM temperature, the comoving
cut-off scale $L_0$ is determined by the condition that the 
linear velocity perturbation $\vec{v}(\vec{x},z)$ averaged on a scale $L_{0}$
is equal to or larger than the IGM sound speed $c_s=(k_BT_{\mathrm{IGM}}(z)/m_p)^{1/2}$.
The IGM temperature is determined by the evolution of the UV background. 
At redshifts $z\le 3$, the temperature varies within the range
$T_{\mathrm IGM}=[10^{3.6}-10^4]$K and its weakly dependent on redshift
(Tittley \& Meiksin, 2007). For $T_{IGM}=10^4$K the sound speed is 
$c_s\simeq 10$km/s; in our subsequent analyses we will fix the sound speed
to this value at all redshifts.
In the linear regime and in comoving coordinates, $\dot{\delta}=-(1+z)\nabla\vec{v}$
and $\dot{\delta}=Hf\delta$ with $f(z)=d\ln\delta/d\ln a$.
In Fourier space, the peculiar velocity $\vec{v}(k)$ on a scale $k=2\pi/L_0$ is 
$|\vec{v}(k,z)|\sim(L_0/2\pi)Hf(z)\delta(k,z)$. From the condition $|\vec{v}|\ge c_s$
and expressing $\delta(k,z)=\delta_0(k)D_+(z)$, with $\delta_0(k)$
the current amplitude of the density contrast at wavenumber $k=2\pi/L_0$, we obtain
$L_0\ge[2\pi c_s](1+z)/[Hf(z)\delta_0(k)D_+(z)]$. This condition is valid only in 
the linear regime, hence the lower bound is obtained by imposing $\delta_0\simeq 1$.
Finally
\begin{equation}
L_0(z)=\frac{2\pi(1+z)c_sH_0^{-1}}{(\Omega_{\Lambda}+\Omega_m(1+z)^3)^{1/2}
f(z)D_+(z)} ,
\label{eq:shock}
\end{equation}
where $H_0$ is the Hubble constant and $\Omega_\Lambda$, $\Omega_m$ are the 
energy density of the cosmological constant and matter density in units of
the critical density. In our numerical estimates, we fixed $\Omega_m$,
$\Omega_\Lambda$ to their concordance values. At $z=0$ the comoving damping scale 
is $L_0\approx 1.7h^{-1}$Mpc.

At redshifts $z \ge 1.0$, shock heating is no longer so effective and the
damping scale $L_{0}$ corresponds to the comoving Jeans length 
at the conditions of the photo-ionized IGM
\begin{equation}
L_0(z)=H_0^{-1}\left[\frac{2\gamma k_{\rm B}T_b(z)}{3\mu m_p\Omega_m(1+z)}\right]^{1/2} ,
\label{eq:jeans}
\end{equation}
where $\gamma$ the polytropic index and
$T_b$ the averaged background temperature of the IGM. 
This last parameter is little constrained by observations; 
Schaye et al (2000) argues that at $z\simeq 3$
HeII re-ionization requires $T_b$ to be larger than $5\times 10^4$K
while Viel \& Haehnelt (2006) gave an upper bound of $T\simeq 2\times 10^5$K.
To simplify, we fix the average background temperature to the constant value $T_b=10^5$K,
within the interval allowed by observations.

\subsubsection{The IGM temperature.}\label{sec:temperature}

To describe the IGM distribution at all redshifts we will consider two 
limiting cases: At $z>1$ the cut-off scale 
is the Jeans length given by eq.~(\ref{eq:jeans}) 
and at $z\le 1$ the shock-heated scale $L_0$ of eq.~(\ref{eq:shock}).
To compute the tSZ contribution to CMB temperature anisotropies due
to the IGM, we need to specify its temperature at each position and
redshift. For the Jeans cut-off length scale we assume the temperature 
follows a polytropic equation of state $T(\hat{x},z)=T_0(z)\xi^{(\gamma-1)}$.
We take $T_0(z)=1.4\times 10^4(1+z)^\beta$K in agreement with the values obtained by 
Hui \& Haiman (2003), with a weak dependence on redshift ($\beta\approx 0$).
We chose $\gamma=1.5$ and $\beta=1$ as a conservative
upper limit to WHIM tSZ anisotropies
at $z\ge 1$. At $z\le 1$, the shock heated IGM has a complex distribution
of densities and temperatures. Kang et al (2005), hereafter K05, computes 
phase-space diagrams that can be fitted by the following equation of
state: $\log_{10}(T_e(\xi)/10^8K)=-2/\log_{10}(4+\xi^{\alpha+1/\xi})$,
valid for overdensities $\xi\le 100$. Alternatively, Cen \& Ostriker (2006), hereafter C06, 
find lower IGM temperatures; their phase-space diagram approximately corresponds
to the equation of state  $\log_{10}(T_e(\xi)/10^8K)=-2.5/\log_{10}(4.0+\xi^{0.9})$.
We have considered all equations of state to be independent of redshift
except the polytropic one. These models are represented in Fig.~\ref{fig:fig1}a; 
solid (black), dashed (blue) and dot-dashed (red) lines correspond to K05 with
$\alpha=(3,1.5,1)$, respectively. The triple-dot
dashed (green) line corresponds to C06 and the
dotted (gold) line corresponds to the polytropic model at $z=1$.

The overall amplitude of the cross-correlation function is proportional to
the fraction of baryons in the WHIM and of the mean temperature of the 
electron gas. In our numerical estimates we have assumed that this baryon fraction
is the same at all redshifts and equal to $f_e=0.5$. The overdensity 
weighted temperature average $\bar{T}_e\equiv\langle 
T_e\xi\rangle/\langle\xi\rangle$ depends on the temperature model. 
For the K05 and C06 models this average in the interval overdensity 
$\xi=[1,100]$ is $\bar{T}_e\approx[20,7,3,0.7]\times 10^6$K,
weakly dependent on redshift; for the polytropic model, whose equation of
state varies with redshift, the mean temperature is in the range 
$\bar{T}_e=[0.4-1.7]\times 10^6$K.
Any constrain on the amplitude of the cross-correlation will
translate into an upper limit on the the product $f_e\bar{T}_e$
and if $f_e$ is independently measured, then it would be a constrain
on the mean temperature of the IGM, offering
a direct probe onto the physical state of the WHIM.

\subsection{Lensing kernel}

\begin{figure}[t]
\centering
\epsfxsize=\textwidth \epsfbox{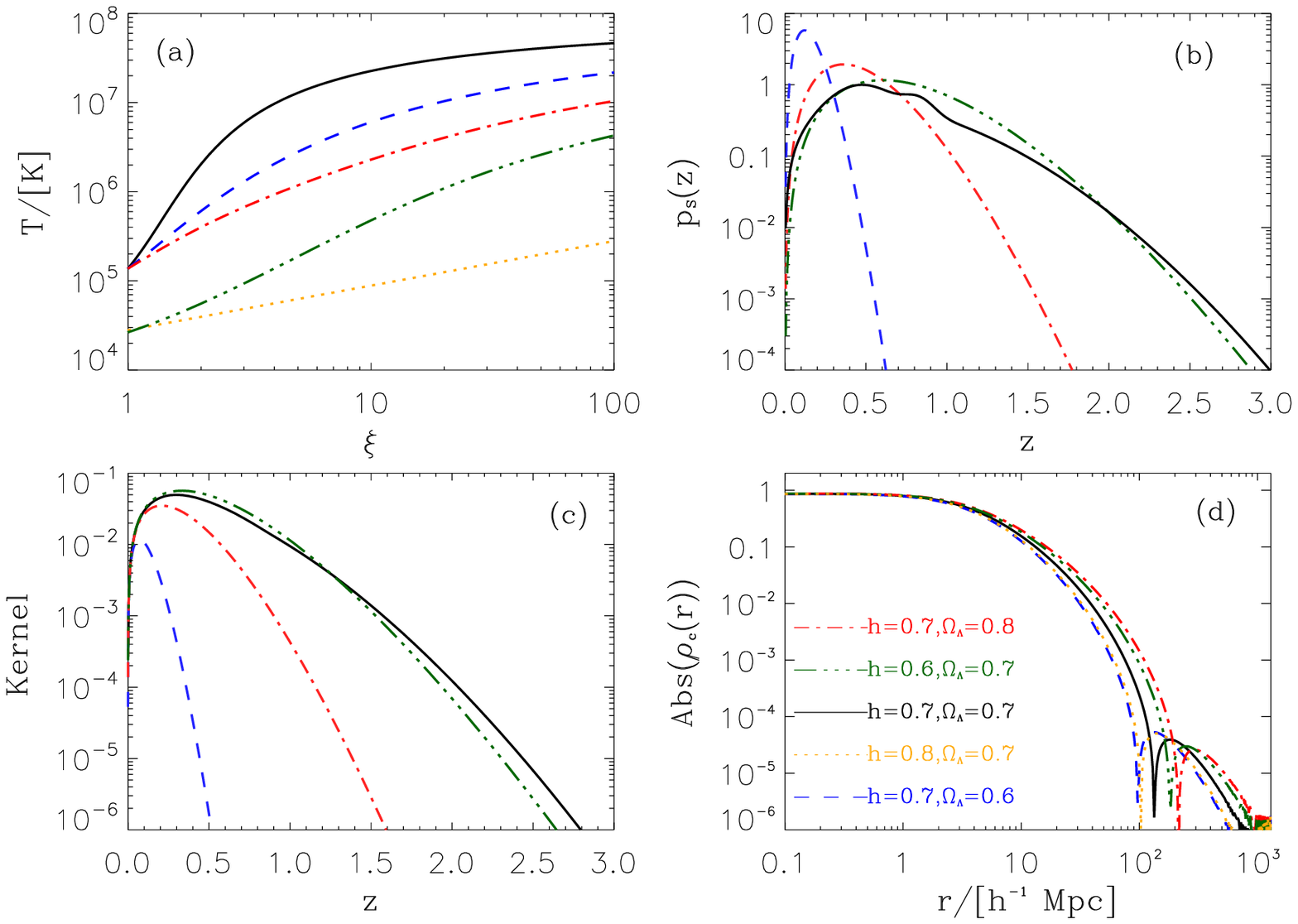}\hfil
\caption{\small 
(a) IGM temperature models considered in this paper;
solid (black), dashed (blue) and dot-dashed (red) lines correspond to 
K05 with $\alpha=(3,1.5,1)$, the triple-dot
dashed (green) line corresponds to C06 and the
dotted (gold) line to the polytropic model at $z=1$, respectively.
(b) Distribution of the lensing sources for 
$z_0=0.1$ (dashed blue), $z_0=0.3$ (solid blue)
and $z_0=0.5$ (dot-dashed blue). For comparison, the thick
solid black lines shows the distribution of the RCSLenS sources
with $mag_r>18$. 
(c) Lensing kernel for the four distributions of the
lensing sources. Lines follow the same conventions as
in (b). (d) Correlation coefficient of eq.~(\ref{eq:rc}) for different 
cosmological parameters.
}
\label{fig:fig1}
\end{figure}

The gravitational field generated by 
weak density perturbations lenses the radiation propagating in the 
Universe. The deflection angle of the weakly deflected rays can be related to 
an effective surface-mass density $\kappa_{eff}$, known as convergence, 
closely related to the mass distribution (Bartelmann \& Schneider 2001).
The convergence due to a population of lenses distributed 
as $G(w)dw=p(z)dz$ along the line of sight is 
\begin{equation}
\kappa_{eff}(\hat{x})
= \frac{3H_0^2\Omega_m}{2c^2}\int_0^{w_H} W(w)
f_K(w)\frac{\delta[f_K(w)\hat{x},w]}{a(w)}dw ,
\label{eq:convergence}
\end{equation}
where $\hat{x}$ is the direction in the sky into which the light ray
starts to propagate, $f_K(w)$ is the comoving angular diameter distance at $w$,
$\delta$ the matter density contrast along the unperturbed light ray
and $c$ the speed
of light. The kernel weights the
relative contribution of lenses along the line of sight
\begin{equation}
W(w)\equiv\int_w^{w_H}dw^\prime G(w^\prime)\frac{f_K(w^\prime-w)}{f_K(w^\prime)} .
\label{eq:weight}
\end{equation}
The redshift distribution of galaxies is modeled as $p(z)=A(z/z_0)^2\exp[-(z/z_0)^{3/2}]$,
where $z_0$ is the effective depth of the lens population, related to the mean redshift 
of the distribution as $z_m=1.412z_0$ (Smail et al 1995); 
the normalization constant is fixed by setting $\int p(w)dw=1$. 
The distribution of galaxies in the CFHTLenS peaks 
at $z=0.37$ (van Waerbeke et al 2014) that corresponds to $z_0=0.3$. 
These lens distributions are represented in Fig.~\ref{fig:fig1}b; 
dashed (blue), dot-dashed (red) and triple dot-dashed (green) lines correspond
to $z=0.1,0.3,0.5$ respectively.  The more recent
analysis by Hojjati et al (2016) uses the deeper RCSLenS catalog
that includes all galaxies with $mag_r>18$. 
These authors provide a numerical fit of their lens distribution, 
plotted in Fig.~\ref{fig:fig1}b with a solid (black) line.

 From eq.~(\ref{eq:convergence}), the contribution from lenses on a thin 
shell of width $dz$ at comoving distance $w=w(z)$ and direction $\hat{x}$ is
\begin{equation}
\Delta\kappa_{eff}=\frac{3H_0^2\Omega_m}{2c^2}
W(z) f_K(z)\frac{\delta[f_K(z)\hat{x},z]}{a(z)}\frac{dw}{dz}dz .
\label{eq:delta_convergence}
\end{equation}
In Fig.~\ref{fig:fig1}c we plot the convergence of eq.~(\ref{eq:delta_convergence})
for $\delta(f_K(w)\hat{x},w)=1$ as a function of redshift
for the four lens distributions given in Fig.~\ref{fig:fig1}b, with
lines following the same conventions. The integration range of 
eq.~(\ref{eq:convergence}) 
must extend up to the horizon $w_H$ or up to a redshift $z^{up}$
high enough to include the effect of all possible lenses. 
We took $z^{up}=5$, and no significant differences were found when 
taking $z^{up}=10$. This is expected since the lensing kernel drops 
exponentially following the distribution of the lensing sources 
(see Figs.~\ref{fig:fig1}b and ~\ref{fig:fig1}c).

Eq.~(\ref{eq:delta_convergence}) was derived in the thin lens approximation
and only terms linear in the density contrast were retained. Higher order
terms contain products of the density field but while the density contrast
could be large for a density perturbation crossed by a given ray, the average
overdensity is $\delta\ll 1$ for most rays and higher order
terms can be safely neglected (Bartelmann \& Schneider 2001). Within this
approximation the PDF of the lenses is that of the linear density
field and, consequently, is well described by a gaussian distribution.

\subsection{Lensing -- TSZ cross-correlation}

To compute the correlation function of lenses and WHIM sources of tSZ anisotropies
given by eq.~(\ref{eq:cfull}), the average $\langle\cdots\rangle$ has to
account for the probability distribution of the WHIM filaments and of the lenses.
Let $dP(\xi,\delta)=F(\xi,\delta)d\xi d\delta$ be the probability that a 
filament with overdensity $\xi$ is located at $(\hat{x}_1,z_1)$ when an overdensity 
$\delta$ is at $(\hat{x}_2,z_2)$, with $F(\xi,\delta)$ the associated
PDF. Then, the average in eq.~(\ref{eq:cfull}) can be written as
\begin{equation}
\langle \Delta Y_C(\hat{x}_1,z_1)\Delta\kappa_{eff}(\hat{x}_2,z_2)\rangle(\theta)=
\int_0^{z_1^{up}}dz_1\int_0^{z_2^{up}}dz_2\int_1^{100}d\xi\int_{-\infty}^\infty d\delta
\Delta Y_C(\hat{x}_1,z_1)\Delta\kappa_{eff}(\hat{x}_2,z_2)
F(\xi,\delta) ,
\label{eq:corr}
\end{equation}
with $\cos\theta=\hat{x}_1\cdot\hat{x}_2$ and $z_1^{up}$ and $z_2^{up}$ the 
highest redshifts beyond which WHIM and lenses do not generate 
a significant cross-correlation. 
To complete our model we need to specify the bivariate PDF of the
lens-filament distribution, $F(\xi,\delta)$. As discussed above,
lensing is dominated by the large scale structure and 
the lensing overdensities $\delta$ are well described by a gaussian PDF, 
but the non-linear overdensities $\xi$ of the IGM filaments are distributed 
according to a log-normal PDF.
Since $\xi=Ae^{\delta_B}$ is log-normal distributed, $\log(\xi)$ follows
a gaussian distribution with mean $\mu_\xi=-\sigma_B^2/2$ and
variance $\sigma_B^2$; in term of this variable the probability
can be written as $dP={\cal G}(\log(\xi),\delta)
d\log(\xi)d\delta$ where ${\cal G}$ is a bivariate gaussian and 
\begin{equation}
F(\xi,\delta)=\frac{1}{2\pi\xi\sigma_B\sigma_\delta(1-\rho_c)^{1/2}}
\exp\left[\frac{1}{2(1-\rho_c^2)}
\left(\frac{(\log\xi-\sigma_B^2/2)^2}{\sigma_B^2}-
2\rho_c\frac{(\log\xi-\sigma^2_B/2)(\delta-\mu_\delta)}{\sigma_B\sigma_\delta}
+\frac{(\delta-\mu_\delta)^2}{\sigma_\delta^2}\right)\right] .
\label{eq:pdf}
\end{equation}
In this expression $\mu_\delta$ is the mean of the matter density contrast, in this case
$\mu_\delta=0$. The variance of the matter density field is
$\sigma_\delta^2=(D_+^2(z)/2\pi^2)\int P_{DM}(k)k^2dk$.
At small scales, $P_{DM}(k)\propto k^{-3}$ and the integral is logarithmically
divergent. Therefore, we remove small scale perturbations by filtering the density
field with a {\it top-hat} window of radius
$R_{cut}=0.5h^{-1}$Mpc. Physically this corresponds to removing from the 
lensing kernel the contribution from galaxies, groups and clusters. Then
\begin{equation}
\sigma_\delta^2=\frac{D_+^2(z)}{2\pi^2}\int P_{DM}(k)W^2_{th}(kR_{cut})k^2dk ,
\end{equation}
where $W_{th}(kR_{cut})$ is the Fourier transform of the {\it top-hat} filter.
Changing the cut-off scale to $R_{cut}=1h^{-1}$Mpc reduces $\sigma_\delta$ by a 
factor 0.85.  The coefficient $\rho_c=\langle\log\xi\delta\rangle/\sigma_B\sigma_\delta$
is the correlation between two gaussian variables 
\begin{equation}
\rho_c(r)=\frac{D_+(z_1)D_+(z_2)}{2\pi^2\sigma_\xi\sigma_\delta}
\int \frac{P_{DM}(k)}{1+L_0^2(z_1)k^2}W_{th}(kR_{cut})j_0(k|\vec{x}_1-\vec{x}_2|)k^2dk ,
\label{eq:rc}
\end{equation}
where $j_0$ is the 0th order spherical Bessel function and
$r=|\vec{x}_1-\vec{x}_2|$ is the comoving distance between a filament in the
IGM at $\vec{x}_1$ and a lens at $\vec{x}_2$, corresponding to 
redshifts $z_1$ and $z_2$, and separated by an angle $\theta$. 
As in Suarez-Vel\'asquez, M\"ucket \& Atrio-Barandela (2013a) 
we use the flat sky approximation and
\begin{equation}
r\equiv|\vec{x}_1-\vec{x}_2|\approx \sqrt{l_{\perp}(\theta,z_1)^2 + [w(z_1)-w(z_2)]^2} ,
\label{eq:flat_sky}
\end{equation}
with $l_\perp(\theta,z_1)$ being the transverse distance between two points
located at the same redshift. Notice that $\rho_c(0)\ne 1$, since the two
distributions, IGM and lenses, are not fully correlated.

In Fig.~\ref{fig:fig1}d we represent the absolute value
of the correlation coefficient for different
cosmological parameters. We assume a flat Universe, i.e., $\Omega_m+\Omega_\Lambda=1$.
The matter power spectrum is normalized to $\sigma_8=0.8$. We verified that
varying parameters within the ranges given in Fig.~\ref{fig:fig1}d has an
effect on the comptonization-convergence cross-correlation that is small
compared with the differences in the lens distribution or the equation 
of the state of the IGM temperature, so we will not discuss further 
variations of cosmological parameters and their effect on our results.

\section{Results and discussion.}

We compute the comptonization-convergence cross-correlation using eq.~(\ref{eq:corr}).
The integration over the lensing part extends up to the redshift
of the last scattering surface. However, as Fig.~\ref{fig:fig1}c indicates, the 
lensing kernel drops exponentially following the distribution of the lensing sources
and, effectively, we can stop the integration at $z_2^{up}=1,2.6,4.6$ for lens
distributions with $z_0=0.1,0.3,0.5$, respectively, when the kernel has decreased by
a factor $10^{-15}$ from its maximum value. For RCSLenS sources
the integration stops at $z_2^{up}=4.6$ when similar drop factor 
has been reached. We verified that, as expected, extending the 
integration further does not increase the cross-correlation.

\begin{figure}[t]
\centering
\epsfxsize=\textwidth \epsfbox{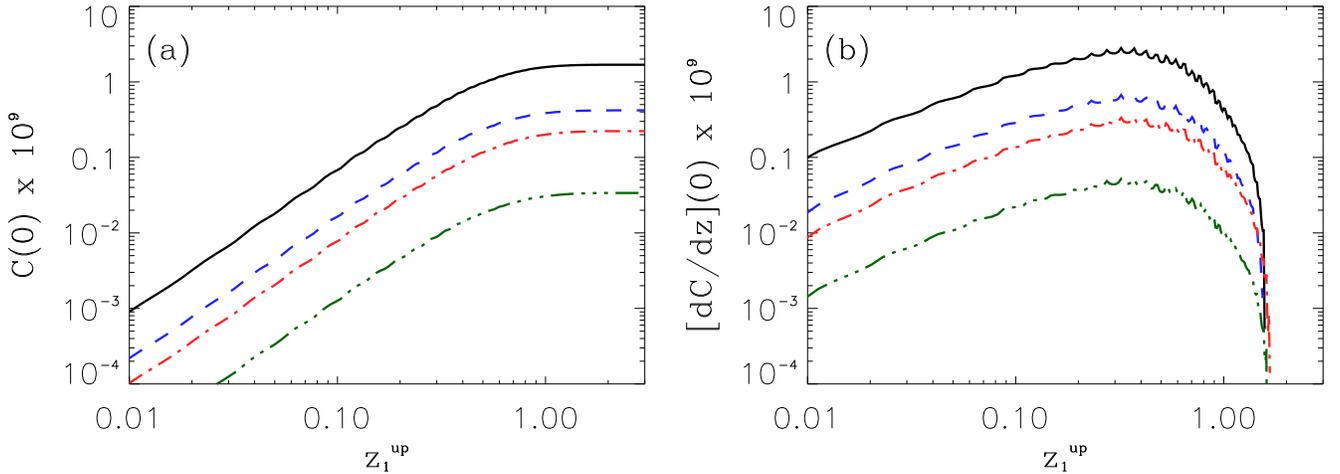}\hfil
\vspace*{-6cm}
\caption{\small 
Amplitude of the convergence-comptonization cross-correlation
at the origin, $C(0)=\langle\kappa_{eff}Y_C\rangle(0)$ as a function
of the upper limit of integration $z_1^{up}$. The results,
from top to bottom correspond to K05 with
$\alpha=3,1.5,1$ (black solid, dashed blue and dot-dashed red lines)
and C06 (triple dot-dashed line).
The lens distribution is the numerical fit to the RCSLenS sources 
The baryon cut-off scale is given by eq.~(\ref{eq:shock}).
}
\label{fig:fig2}
\end{figure}

More delicate is to decide out to what redshift is valid our model of the IGM. 
At $z\ge 1$ shock-heating
stops being dynamically important. In Fig.~\ref{fig:fig2}a we compute the
amplitude of the effective lensing-comptonization  
cross-correlation at zero lag, $C(0)=\langle\kappa_{eff}Y_C\rangle(0)$
as a function of the upper limit of integration $z_1^{up}$. The results,
from top to bottom correspond to the K05 with
$\alpha=3,1.5,1$ (black solid, dashed blue and dot-dashed red lines)
and the C06 model (triple dot-dashed green line).
In Fig.~\ref{fig:fig2}b we plot the differential contribution. This figure
indicates that most of the cross-correlation originates from $z\le 1$.

Since the cross-correlation scales with the fraction of electrons in 
the IGM as $\langle\kappa_{eff}Y_C\rangle\propto(f_e/0.5)$, we need to
know the fraction of electrons in the WHIM to translate constraints
on $C(0)$ into constraints in the mean temperature of the gas.
Although 80\% of all baryons reside in Ly$\alpha$ systems at redshift 
$z\simeq 2$ and $f_e\le 0.2$ at that redshift
(Fukugita et al 1998), numerical simulations
indicate that $f_e\ge 0.4$ out to $z\simeq 1$ (C06), the range in
redshift space that dominates the cross-correlation. Therefore, by taking $f_e=0.5$ and
constant our constraints on the mean WHIM temperature will be reasonably accurate.

The contribution of the IGM comptonization parameter to the cross-correlation from 
$z\ge 1$ is less than 10\%. In fact, this correction is overestimated. First,
the fraction of baryons in the WHIM drops with redshift. Second, and
as mentioned in Sec~\ref{sec:temperature}, the IGM behaves as a polytrope
and, on average, its temperature is smaller than the K05
shock-heated models and is similar to C06
(see also Fig~\ref{fig:fig1}a). In the interval $z\ge 1$ the cross-correlation
with the polytropic equation of state and the damping scale of eq.~(\ref{eq:jeans}) 
is $\langle\kappa_{eff}Y_C\rangle\sim 1-10\times 10^{-12}$ for the different lens
distributions. This is a very small contribution and essentially we
could have stopped our calculation at $z=1$ or extend the 
shock model out to $z=3$ since it would have introduced an error 
smaller than 10\%. We adopted this latter option and by not including
the Jeans cut-off scale (eq.~\ref{eq:jeans}) and its corresponding
polytropic equation of state, we simplify the parameter space
of our model and the physical interpretation of our results.

\begin{figure}[t]
\centering
\epsfxsize=\textwidth \epsfbox{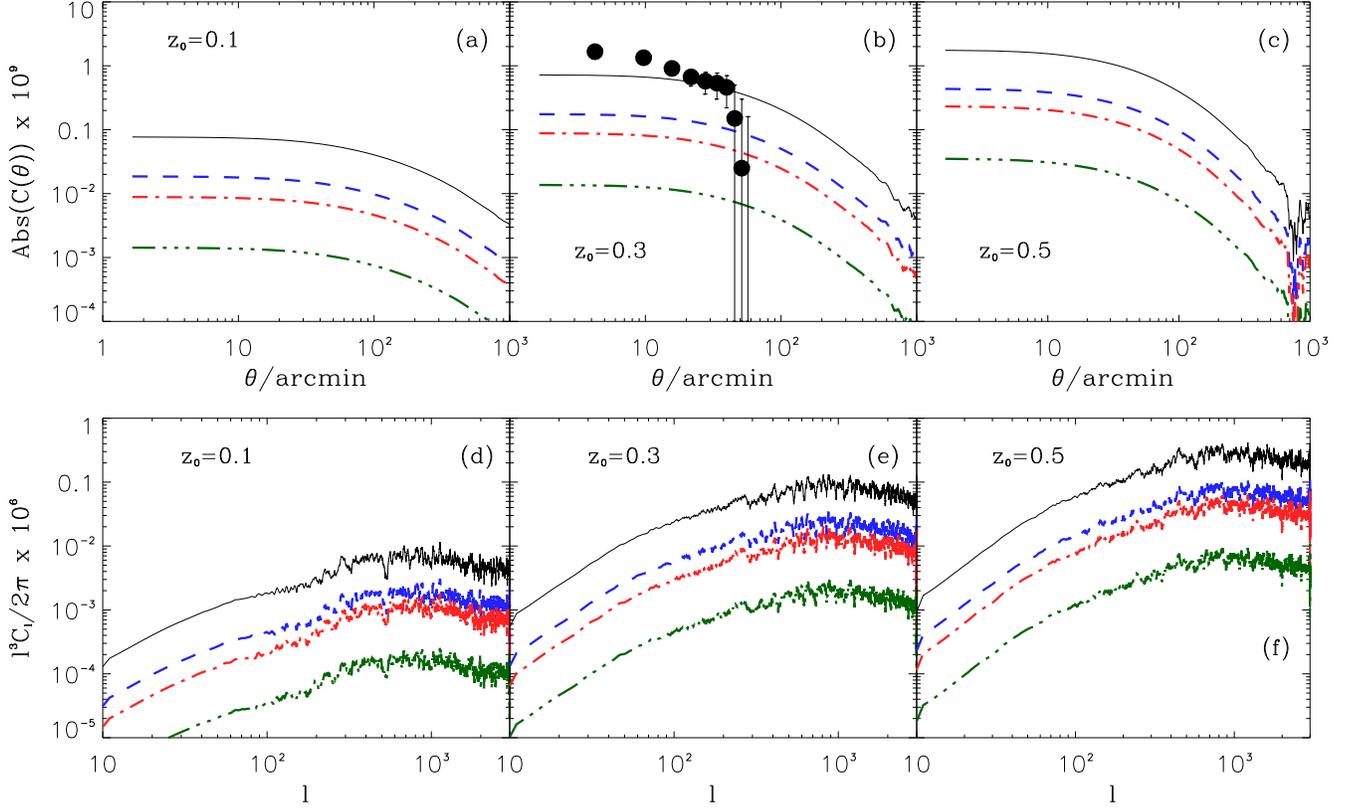}\hfil
\vspace*{-2cm}
\caption{\small Comptonization parameter-lensing convergence cross-correlation
(upper panels) and power spectrum (lower panels) for lens source catalogs
of different scale $z_0$.  The cut-off scale is given by eq.~(\ref{eq:shock}).
From top to bottom, lines correspond to K05 with $\alpha=3,1.5,1$
(solid black, dashed blue and dot-dashed red, respectively) and C06
(triple dot-dashed line). The corresponding power spectra
are shown in the bottom panels with line following the same conventions.
In (b) the correlation data and error bars were taken from Van Waerbeke et al (2014). 
}
\label{fig:fig3}
\end{figure}

\begin{figure}[t]
\centering
\epsfxsize=\textwidth \epsfbox{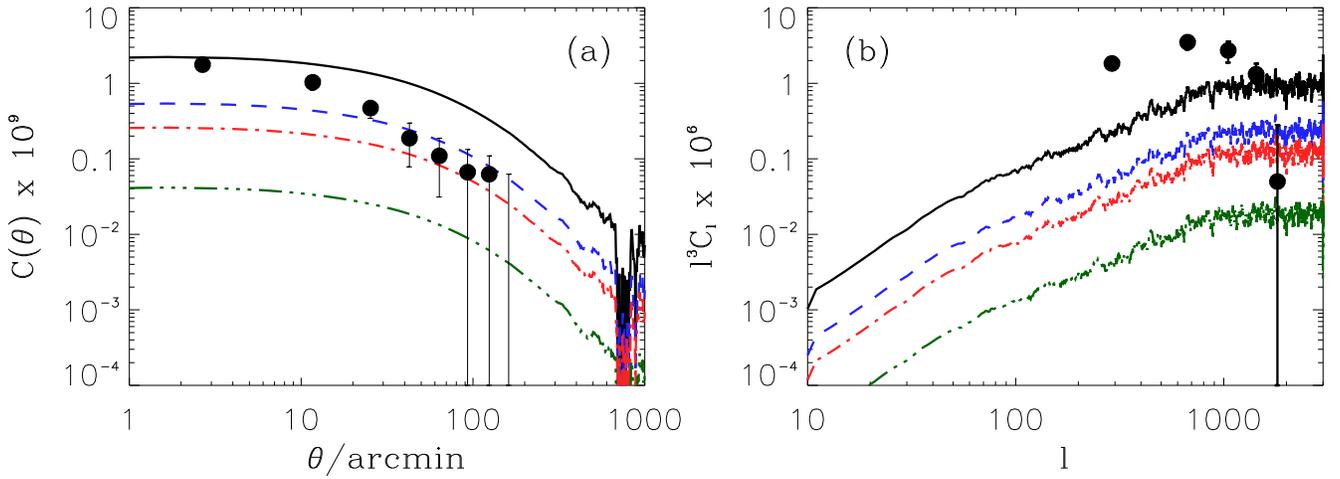}\hfil
\vspace*{-6cm}
\caption{\small 
Comptonization-convergence correlation and power spectrum
due to the RCSLenS sources with $mag_r>18$. 
Lines follow the same conventions as in Fig.~\ref{fig:fig3}. 
The correlation and power spectrum data points were taken from Hojjati et al (2016). 
}
\label{fig:fig4}
\end{figure}

Figs.~\ref{fig:fig3} and ~\ref{fig:fig4} constitute our main result. 
In Fig.~\ref{fig:fig3} we plot the $\langle\kappa_{eff}Y_C\rangle$ cross-correlation
for the three lens distributions with $z_0=0.1,0.3,0.5$ (upper panels) 
and their corresponding power spectra (lower panels). 
The power spectrum is computed from the correlation function integrating the
quadrature
\begin{equation}
C_\ell=2\pi\int\langle \kappa_{eff}Y_C\rangle P_\ell(\cos\theta)d\cos\theta ,
\end{equation}
with $P_\ell$ the $\ell$-th Legendre polynomial. Hence, we are required to
compute the correlation function over the range $\theta=[0,\pi]$rad.
To simplify our calculation we have assumed the sky to be flat 
(eq.~\ref{eq:flat_sky}) and although this approximation limits the accuracy of the
low-$\ell$ multipoles, it should be accurate for multipoles where data
is available, $\ell\ge 100$.  

In the panels of Fig.~\ref{fig:fig3} and
from top to bottom the solid (black), dashed (blue) and dot-dashed (red)
lines correspond to K05 with $\alpha=3,1.5,1$
and the triple dot-dashed (green) line corresponds to C06. 
The distribution of the CFHTLenS is well approximated by 
$z_0=0.3$ then, in Fig.~\ref{fig:fig3}b we also plot the 
data from van Waerbeke et al (2014) and their respective error bars. 
In Fig.~\ref{fig:fig4} we plot the correlation function and the power
spectrum for the same shock-heating temperature models but for the 
RCSLenS sources with $mag_r>18$. Lines follow the same conventions
than in Fig.~\ref{fig:fig3}. The data are taken from Hojjati et al (2016).

To analyze the contribution of the different IGM overdensities we divide
the integration of eq.~(\ref{eq:corr}) in four intervals with equal
logarithmic spacing: $\xi=([1-3.3],
[3.3-10],[10-33],[33-100])$. We computed the contribution in each interval
for the K05 with $\alpha=1.5$ and for lens
distributions $z_0=0.3$ (Fig.~\ref{fig:fig3}b) and RCSLenS sources
(Fig.~\ref{fig:fig4}a). The fractional contribution to the
correlation at the origin, $\langle\kappa Y_C\rangle$(0),
was $(0.08,0.45,0.41,0.06)$
for the first case and $(0.1,0.39,0.46,0.05)$ in the second. Similar results
occur for other lens distributions and temperature models: 
most of the correlation comes from
overdensities in the range $\xi\approx[3-33]$. The numerical simulations
of Dav\'e et al (2001) found that this is the density range where most
of the WHIM is stored. In this respect, if our log-normal model were to 
be accurate only at these intermediate overdensities, integration
of eq.~(\ref{eq:corr}) would still provide a very accurate result.

The comparison of the measured data with the theoretical predictions
already offers some insights into the nature of the IGM.
For the CFHTLenS sources shown in Fig~\ref{fig:fig3}b, all temperature models
are allowed by the data. As indicated in Sec.~\ref{sec:temperature},
the shock-heating model with $\alpha=3$ corresponds to an average 
temperature of $\bar{T}_e=2\times 10^7$K and is still compatible with
the measured correlation. However, our results do not include
the contribution due to clusters and galaxy groups. 
Since at most 15\% of the measured signal comes from unbound gas if 
we restrict the overall IGM contribution
to be that fraction of the overall signal then only models 
with $\alpha\le 1.5$ are compatible with the data. In other words,
the mean temperature of the IGM free electron gas would
be $\bar{T}_e\le 7\times 10^6$K.

The data from Hojjati et al (2016) shown in Fig~\ref{fig:fig4}a
are even more restrictive. These authors
compared their measurements against the predictions of the
halo model and from numerical simulations that included diffuse gas.
The simulations showed a very good agreement
with the observed cross-correlation from RCSLenS galaxies, with about 10-15\%
contributions coming from unbound gas. The amplitude of the correlation
(Fig~\ref{fig:fig4}a) in the range $\theta=[40-120]$arcmin is that
of the $\alpha=1.5$ model. In that interval, only the C06
temperature model predicts an amplitude 10\% of the measured 
correlation. That would imply that the average temperature
of the IGM is $\bar{T}_e\sim 10^6$K, a stricter bound than derived
from the van Waerbeke et al (2014) data.

There is a caveat when translating the results on the cross-correlation onto
an upper limit on the average temperature of the IGM. Hydro-simulations
consistently show that unbound gas is not well characterized by a single
equation of state; more accurately, the gas coexists in different phases and
there is a large spread in temperature within regions with the same overdensity.
Since our temperature models fails to encode the full complexity
of the temperature-density phase diagram, our upper bounds on the average temperature
must be understood as an order of magnitude estimate not as a strict upper limit.

The constrains that can be derived from the measured 
power spectrum shown in Fig.~\ref{fig:fig4}b are not as tight as those
derived from the correlation function. Only the measurement at $\ell\sim
1800$ is well below the prediction for the K05 models.
What is more relevant is that the overall shape is very different.
Hojjati et al (2016) found that the shape of the correlation function
and power spectrum was strongly dependent on physical processes undergone
by baryons in halos such as radiative cooling, star formation, supernovae winds and 
AGN feedback. For instance, AGNs expel gas to large 
distances from the center of halos, lowering the signal at small scales.
The properties of the hot gas in our model are rather simplified. No effects
of specific physical processes are considered and only the
density and temperature distributions are important. More realistic
models would require detailed numerical simulations including
the most relevant processes in low density regions. Physical effects could
remove power at $\ell\ge 1000$, modifying the overall shape of the
power spectrum and bringing it in closer agreement with the data.
While a detailed discussion on this point is beyond the scope of the
current paper, if the shape were to be independent of the physics of baryons
the power spectrum could be a useful discriminant between halo and 
unbound gas contributions.

An alternative approach to detect the WHIM contribution would be to remove
known galaxies down to a given magnitude to eliminate the contribution of 
their halos to the comptonization-convergence correlation. When removing 
fainter galaxies does not produce a further decrement of the cross-correlation,
we have reached the level when the signal is due to gas outside halos.
A similar approach has been used by Kashlinsky et al (2005) and Helgason
et al (2015) to isolate Cosmic Infrared Background fluctuations due to
first stars at the epoch of reionization
from those of known galaxy populations in deep Spitzer data.

\section{Conclusions.}

Models of galaxy formation predict that a significant fraction, close to
half the total number of baryons, could be stored in the WHIM.
The low densities and temperatures $10^{5-7}$K of this medium makes
it difficult to detect. Search for absorption lines and SZ contributions
have provided preliminary evidence of its existence.
The $\langle\kappa Y_C\rangle$ cross-correlation measured by Van Waerbeke et al 
(2014) and Hojjati et al (2016) probes the fluctuations on
the electron pressure along the line of sight and its distribution but 
it is not yet a detection of the missing baryon component.
As indicated by Hojjati et al (2015) about 50\% of the signal
comes from the small fraction of baryons within massive halos;
at most, 15\% of the cross-correlation power at $\ell\sim 500$ could
come from unbound gas. In this article we
have shown that the contribution from the unbound gas in filaments
could be of this order of magnitude, depending on model parameters.
In particular, if the unbound gas is well described by a log-normal
distribution and the gas is shock heated out to a mean
temperature $\bar{T}_e\sim 10^6$K,
then about half the baryons on the Universe could be stored in the WHIM  
producing a signal that is at least one order of magnitude smaller than the measured
amplitude. We have considered two different baryon cut-off lengths: the Jeans
length given by eq.~(\ref{eq:jeans}) that  would describe 
better the physical state of the IGM at $z>1$ and the shock-heated
cut-off scale given by eq.~(\ref{eq:shock}), that provides a better 
description at $z<1$. We have shown that $\sim 90$\% of the contribution
to the $\kappa_{eff}-Y_C$ cross-correlation and to its power spectrum 
originates at $z\le 1$ and at overdensities in the range $\xi\sim[3-33]$.
The overall amplitude depends on the depth of the source catalog
probing the convergence due to the large scale structure, the average
electron temperature and is proportional to the fraction of baryons in the IGM. 

The tSZ-lensing cross-correlation could be a potentially powerful technique 
to trace the distribution of baryons at large scales. 
The shape of the measured comptonization-convergence power spectrum and the
theoretical prediction for IGM gas show maxima at different scale. 
The difference could be due to 
not having include the physical effects that are relevant to the evolution
of the IGM gas; but if the differences in shape are real, they could 
be used to separate the contribution of unbound gas from that of gas in halos.
In real space, the cross-correlation is also dominated by halos.
To detect the contribution due to the WHIM would require to mask galaxy
populations with increasing magnitude down to the level when further
masking does not reduce the residual correlation. This would require to
extend the measurement to larger areas and to deeper lens surveys, as 
Hojjati et al (2016), to increase the signal-to-noise by a factor 5-10. 
Then, masking the halo contribution down to 10\% of its original 
amplitude would still leave a statistically significant signal.

\vspace*{1cm}
{\bf Acknowledgments}

F. A.-B. acknowledges financial support from the grant FIS2015-65140-P (MINECO/FEDER).
He also thanks the hospitality of the Leibniz Institute f\"ur Astrophysik at Postdam
where part of this work was done.

\clearpage
\pagestyle{plain}

\end{document}